\documentstyle[psfig,epsf,conf-X]{article}
\def\ref{\par\noindent\hangindent=1truecm}
\font\piedi=cmr8

\catcode`\@=11
\def\gsim{\ifmmode{\mathrel{\mathpalette\@versim>}}
    \else{$\mathrel{\mathpalette\@versim>}$}\fi}
\def\lsim{\ifmmode{\mathrel{\mathpalette\@versim<}}
    \else{$\mathrel{\mathpalette\@versim<}$}\fi}
\def\@versim#1#2{\lower 2.9truept \vbox{\baselineskip 0pt \lineskip 
    0.5truept \ialign{$\m@th#1\hfil##\hfil$\crcr#2\crcr\sim\crcr}}}
\catcode`\@=12

\def\lsun{\hbox{$L_\odot$}}
\def\lb{\hbox{$L_{\rm B}$}}
\def\msun{\hbox{$M_\odot$}}
\def\IMLR{Fe$M/L$}
\def\micm{\hbox{$M_{\rm ICM}$}}
\def\mfecm{\hbox{$M_{\rm Fe}^{\rm ICM}$}}

\def\zfes{\hbox{$Z^{\rm Fe}_{*}$}}
\def\zfecm{\hbox{$Z^{\rm Fe}_{\rm ICM}$}}
\def\ho{\hbox{$H_\circ$}}
\def\h50{\hbox{$\ho /50$}}
\def\yr-1{\hbox{${\rm yr}^{-1}$}}\begin{document} 
\small
\heading{%
%
What Heavy Elements in Clusters of Galaxies Tell About Clusters and Galaxies
}
\par\medskip\noindent
\author{%
Alvio Renzini
}
\address{%
{European Southern Observatory, D-85748 Garching b. M\"unchen, Germany}
}
\begin{abstract}
Clusters of galaxies allow a direct estimate of the metallicity and
metal production yield on the largest scale so far. The ratio of the
total iron mass in the ICM to the total optical luminosity of the
cluster (the iron mass-to-light-ratio) is the same for all clusters
which ICM is hotter than $\sim 2$ keV, and the elemental proportions
(i.e. the [$\alpha$/Fe] ratio) appear to be solar. The simplest
interpretation of these
evidences is that both the IMF  as well the relative
contributions of SN types are universal. 
Currently available abundances in cooler clusters and groups are much 
more undertain, possibly due to insufficiently accurate atomic physics
data for multielectron ions, or to the ICM being multi-phase, or to a
combination thereof. This uncertainty automatically extends to the reality
of radial abundance gradients so far reported in cool clusters.
It is enphasized that most metals reside in the ICM rather than in galaxies,
which demonstrates that energetic winds operated early in the evolution 
of massive galaxies, the likely producers of most metals now in the ICM.
The ICM metallicity is also used to set a semiempirical constraint of
$\sim 0.1$ keV per particle to the ICM {\it preheating} due to
supernova driven galactic winds. A lower limit of the universe global
metallicity at $z=3$ is also derived.
\end{abstract}
\section{Introduction}

Clusters of galaxies offer many interesting opportunities to astrophysical
and cosmological research. One of them  is the
study of the cosmic chemical evolution on the largest scale so far.
In turn, the heavy element content of clusters results in 
interesting constraints on the formation and evolution of clusters as well
as of their population of galaxies. 

Current estimates of iron content of the intracluster medium (ICM) are
presented in Section 2, together with those of the abundance of the
$\alpha$-elements. Section 2 also deals with the reported
radial abundance gradients in cool clusters, as well as with the
relative share of iron and the other heavy elements now in the ICM and
locked into stars and galaxies.  Section 3 deals with the origin of
the ICM metals, identifying in the old, giant spheroids the main producers.
Section 4 offers an estimate of the ICM preheating by early
supernova driven galactic winds, while Section 5 provides references for 
further reading concerning the implications for the globla metallicity of 
the universe at low as well as high redshift.

\section{Iron and $\alpha$-Elements in the ICM}
As Larson and Dinerstein (1975) predicted and X-ray observations confirmed
(Mitchell et al. 1976), the ICM is rich in heavy elements. Actually, it
has the larger share of them compared to cluster galaxies.
\subsection{Iron}
Fig. 1 shows the 
iron abundance in the ICM of clusters and groups as a function of ICM
temperature from an earlier compilation 
(Renzini 1997, hereafter R97). Data cluster along two sequences in Fig. 1:
for $kT\gsim 3$ keV the ICM iron abundance is constant, i.e. independent of  
cluster temperature. Abundances for clusters in this {\it horizontal} sequence 
come from the {\bf Iron-K} complex at $\sim 7$ keV, which emissions
are due to transitions to the K level of H-like and He-like iron ions. 
Conversely,
at lower temperatures data delineate an almost {\it vertical} sequence,
with a fairly strong abundance-temparature correlation reaching
extremely low iron abundances in groups cooler than $\sim 1 $ keV.
Iron abundances in the vertical sequence from the {\bf Iron-L} complex
at $\sim 1$ keV, which emission lines are due to transitions to the L level of
iron ions with three or more electrons. In these cooler clusters iron is indeed
in such lower ionization stages, and the iron-K emission disappears.
Such low
abundances were regarded with suspicion in R97, being possibly the
result of systematic inaccuracies in the  atomic physics data used to
model the iron-L complex. 

Fig. 2 is the same of Fig. 1
with one  difference. Clusters
and groups in R97 with $kT<2$ keV have been substituted
by a set of 11 groups from Buote (1999). Buote
argues that the atomic physics may be correct, but gets
substantially higher (perhaps more realistic) abundances when using
2-temperature fits. It remains to be established
whether the apparent correlation
of the iron abundance with temperature is real or instead is an artifact of 
inadequate diagnostics at low temperatures. For this reason conclusions 
concerning abundances derived from iron-K should be regarded as more secure 
than those for cooler clusters.

Fig. 3 shows the iron-mass-to-light-ratio (\IMLR) of the ICM, i.e.
the ratio $\mfecm/\lb$ of the total iron mass in
the ICM over the total $B$-band luminosity of the galaxies in the
cluster. Again, objects with $kT<2$ keV in R97 have
been omitted, with the exception of 4 objects from Buote (1999) for
which a measure of the ICM mass was available from the R97 compilation.

\begin{figure}
\vskip-4.truecm
\centerline{\hskip+5mm\psfig{file=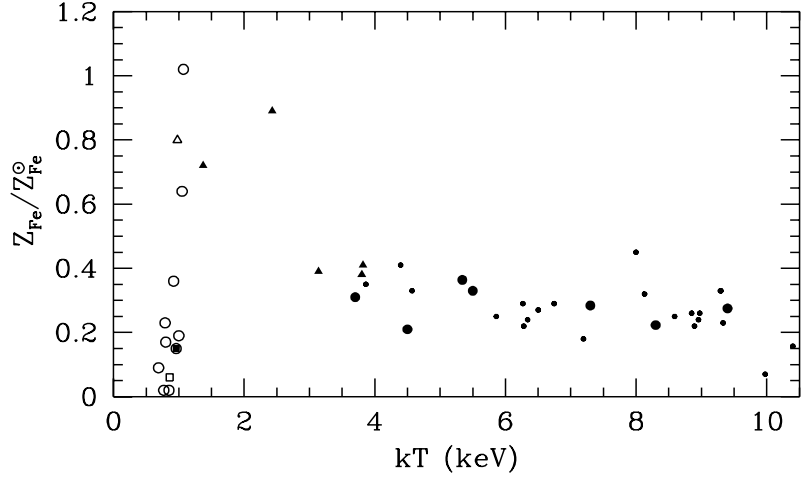,width=11.5cm,angle=0,height=11.2cm}}

\vskip-0.5truecm
\caption{\piedi A compilation of the iron abundance in the ICM as a 
function of ICM temperature
for a sample of clusters and groups (R97), including several clusters at
moderately high redshift with $<z>\simeq 0.35$, represented by small
filled circles.}
\end{figure}
\begin{figure}
\vskip -4 truecm
\centerline{\hskip+5mm\psfig{file=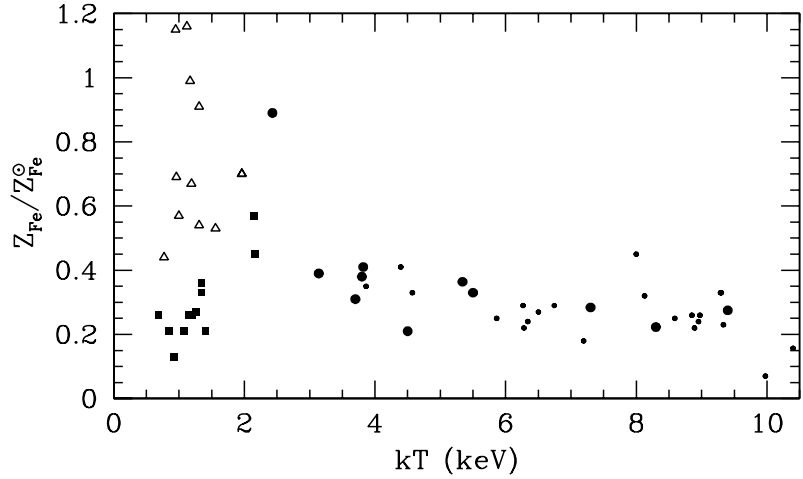,width=11.5cm,angle=0,height=11.2cm}}
\vskip-1truecm
\caption[]{The same as Fig. 1 with clusters and groups with 
$kT\lsim 2$ keV having been replaced by 11 groups 
from Buote (1999), which temperatures and abundances are determined
from 1- and 2-temperature fits (filled squares and open triangles,
respectively).}
\vskip - 0.5 truecm
\label{eps1}
\end{figure}

The drop of the \IMLR \ in poor clusters and groups (i.e. for
 $kT\lsim 2$ keV) noticed in R97 can be traced back to a drop in both 
the iron abundance (which however may not be real)
 {\it and} in the ICM mass to light ratio. Such groups appear to be gas poor
compared to clusters, which suggest that they may 
 be subject to baryon and metal losses due to strong
 galactic winds driving much of the ICM out of them (Renzini et
 al. 1993; R97: Davis et al. 1998). For the rest of this 
paper I will concentrate on clusters with $kT\gsim 2-3$ keV.

Fig.s 1-3 show that both the iron abundance and the
\IMLR \ in rich clusters ($kT\gsim 3$ keV) are independent of
cluster temperature, hence of cluster richness and
optical luminosity. For these clusters one has 
$\zfecm =0.3\pm 0.1$ solar, and $\mfecm/\lb = (0.02\pm 0.01)$ for
$\ho=50$.
The most straightforward interpretation is that clusters did not lose
iron (hence baryons), nor differentially acquired pristine baryonic 
material, and
that the conversion of baryonic gas into stars and galaxies has
proceeded with the same efficiency  and the stellar IMF in all clusters (R97).
Otherwise, we should observe cluster to cluster variations of the iron
abundance and of the \IMLR. 

The absence of such variations also argues for the iron now residing
in the ICM having been {\it ejected} from galaxies by supernova- (or
AGN-) driven galactic winds, rather than having been stripped by ram
pressure (Renzini et al.  1993; Dupke \& White 1999). Indeed, ram
pressure effects become much stronger with incereasing cluster
richmess, hence galaxy velocity dispersion and ICM
temperature. Correspondingly, if ram pressure would play a major role
in getting iron out of galaxies one would expect the ICM abundance and
\IMLR \ to increase with $kT$, which is not observed.

\begin{figure}
\vskip -4 truecm
\centerline{\hskip+5mm\psfig{file=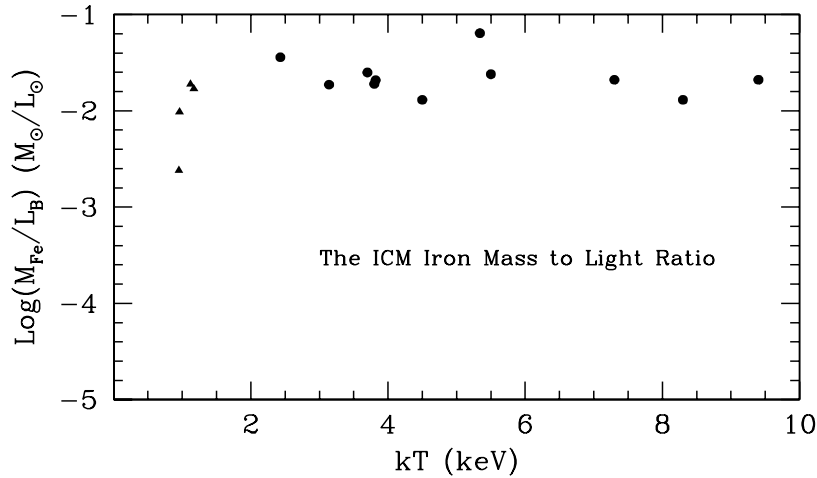,width=11.5cm,angle=0,height=11.2cm}}
\vskip - 0.7truecm
\caption[]{The iron mass to light ratio  of  the  ICM of clusters
and groups (for $\ho=50$) as a function of the ICM temperature (R97).}
\label{eps2}
\end{figure} 

\subsection{The $\alpha$-Elements}

From {\it ASCA} X-ray observations  the ICM 
abundances of some  of the $\alpha$-elements 
(O, Ne, Mg, and Si) have also beed derived, yielding an average
$<\![\alpha$/Fe]$>\simeq +0.2$ (Mushotzky et al. 1996).  This 
modest $\alpha$-element enhancement suggested an ICM
enrichment dominated by SNII products.  However,  the $\alpha$-element
 overabundance vanishes when consistently adopting the
``meteoritic'' iron abundance for the sun, as opposed to the
``photospheric'' value (Ishimaru \& Arimoto 1997), with a formal average
$<\![\alpha$/Fe]$>\simeq +0.04\;\pm\sim 0.2$ (R97).  Clusters of
galaxies as a whole are therefore nearly {\it solar} as far as the
elemental ratios are concerned, which argues for stellar
nucleosynthesis having proceeded in much the same way in the solar
neighborhood as well as at the galaxy cluster scale.  
This argues for a similar ratio of the number of Type Ia to Type II SNs,
as well as a similar IMF (R97), suggesting that the star
formation process (IMF, binary fraction, etc.) is universal, with
little or no dependence on the global characteristics of the parent
galaxies in which molecular clouds are turned into stars.  

\subsection{Radial Abundance Gradients, Are They Real?}

The values of the \IMLR \ shown in Fig. 3 were derived as the product
of the ICM mass times the central iron abundance. However, radial
gradients in the iron abundance have been reported for several
clusters (e.g. Fukazawa et al. 1994; Dupke \& White 1999); Finoguenov,
et al. 1999; White 1999), which would require a radial integration
of $\rho(r)Z_{\rm Fe}(r)dV$ to derive the total iron mass in the ICM.
This is not currently feasible because ASCA data lack of the sufficient
image quality to do a proper job.

Nevertheless, a cursory examination of the radial gradient so far reported
reveals two intriguing trends. The first is that the presence of radial
abundance gradients is confined to cool clusters and groups ($kT\lsim 3$ keV,
with hotter clusters -- which incidentally are generally those with the best
S/N ratios -- never showing appreciable gradients. The second trend one
can notice in the published data is that an abundance gradient is most
often associated to a temperature gradient: iron abundances increase inwards
as temperature drops towards the 
center to $\sim 2 $ keV or below. These trends are illustrated in Fig. 4, 
showing the abundance-temperature relation for a representative set of 
clusters (Pellegrini 2000).

\begin{figure}
\vskip -2.5 truecm
\centerline{\hskip+5mm\psfig{file=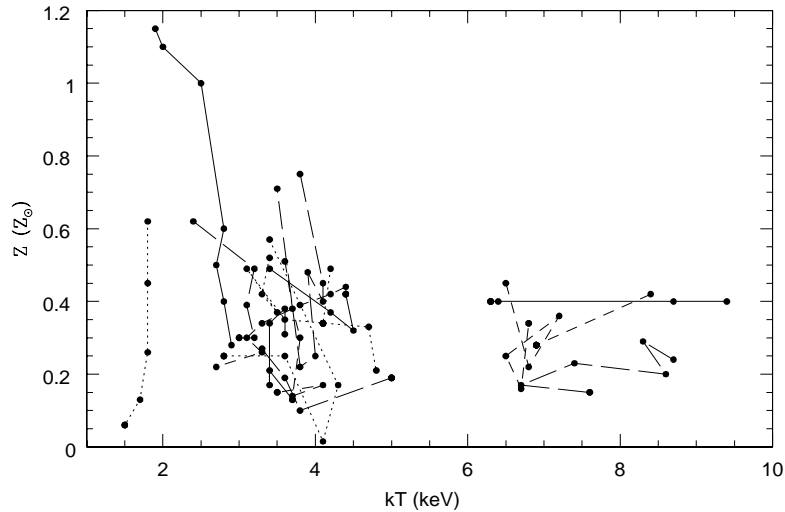,width=11.5cm,angle=0,height=11.2cm}}
\vskip -2.5 truecm
\caption[]{The metal abundance vs temperature within individual clusters
from a compilation by Pellegrini (2000). Data points for a given cluster at
various distances from the cluster center are connected by lines.
}
\vskip - 0.5 truecm
\label{eps4}
\end{figure}

Given the uncertainties affecting the low temperature abundances -- resulting
from either problems with the atomic physics or from the complexities of
a multi-temperature ICM -- it seems prudent to conclude that the reality of
the reported abundance gradients is far from beeing established. 
Although more or less plausible mechanisms could be invented to produce 
radial abundance gradients in clusters, it is neverhteless worth entertaining
the possibility of the reported gradients being a mere artifact of inadequate
diagnostics.

\subsection{Most Metals are not in Galaxies}

The metal abundance of the stellar component of cluster galaxies can
only be inferred from integrated spectra coupled to synthetic stellar 
populations. Much of the stellar mass in clusters is confined to
passively evolving spheroids (ellipticals and bulges) for which the
iron abundance may range from $\sim 1/3$ solar to a few times solar.
Following R97, the ratio of the iron mass in the ICM to the iron mass
loked into stars and galaxies is given by:

\begin{equation}
{\zfecm\micm\over\zfes M_*}\simeq 1.65 h^{-3/2},
\end{equation}
or 4.6, 2.5, and 1.65, respectively for $h=0.5$, 0.75, and 1 ($h=\ho/100$),
and having adopted $\zfecm = 0.3$ solar and $\zfes =1$ solar. 
Note
that with the adopted values for the quantities in equation (2) most
of the cluster iron resides in the ICM, rather than being locked into
stars, especially for low values of $\ho$. 
Therefore, it appears that there is at least as much metal mass out of 
cluster galaxies (in
the ICM), as there is inside them (locked into stars). 
This sets a strong constraint by models of the chemical evolution of
galaxies.

\begin{figure}
\vskip -1.5 truecm
\centerline{\hskip+5mm\psfig{file=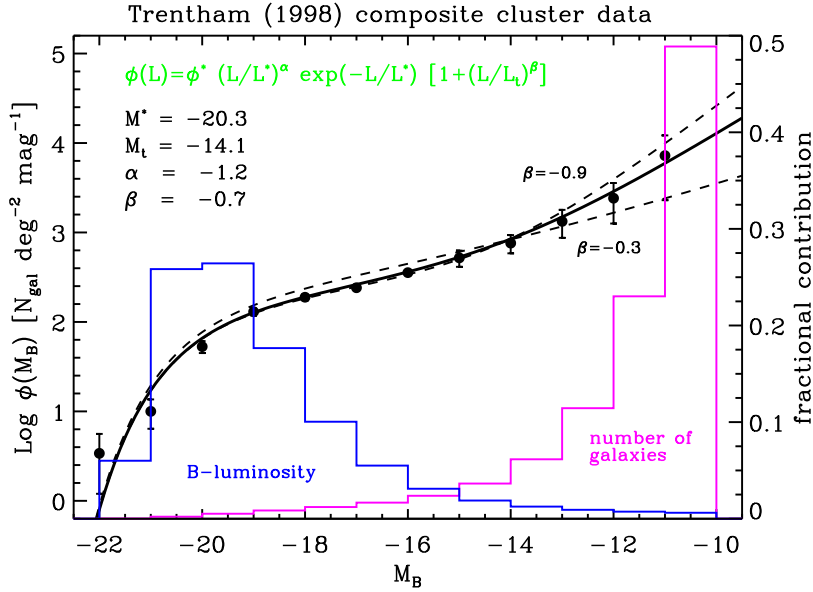,width=11.5cm,angle=0,height=8.2cm}}

\vskip - 0.5  truecm
\caption[]{The galaxy luminosity function from Trentham (1998, dots) and its analytical rendition
$\phi(L)$ are used to visualize the fractional contributions per magnitude bin to the total cluster
luminosity and to the total number of cluster galaxies, left and right histograms, respectivel
(from Thomas 1999).}

\label{eps1}
\end{figure}

\section{Origin of the Metals in the ICM}

The constant \IMLR \ of clusters says that the total mass of iron in
the ICM is proportional to the total optical luminosity of the cluster
galaxies (Songaila et al. 1990; Ciotti et al. 1991; Arnaud et
al. 1992; Renzini et al. 1993). This strongly argues for the iron
(metals) now in the ICM having been produced by the (massive) stars of
the same stellar population to which belong the low-mass stars now
producing the bulk of the cluster optical light.  As well known, much
of the cluster light comes from the old spheroidals (ellipticals and
bulges), hence one can say that {\it the bulk of cluster metals were
produced by ellipticals and bulges when they were young}.

It is now well established that the stellar populations in spheroidals --
in clusters as well in the field -- are very old, with the bulk of stars 
having formed at $z\gsim 3$ (for a recent review with extensive references
see Renzini \& Cimatti 1999). Therefore, the simplest interpretation of the 
data suggests that the bulk of the heavy elements in the ICM were produced
and expelled from galaxies a long time ago, i.e. at $z\gsim 3$. If so, the ICM
abundances should not show any significant evolution all the way to very high
redshifts. Existing data are in agreement with this prediction (see Fig. 1),
but do not reach much beyond $z\simeq 0.5$. Future
X-ray observatories could check this prediction. 

It is also interesting to address the question of which galaxies have
produced the bulk of the iron and the other heavy elements, i.e. the relative
contribution as a function of the present-day luminosity of cluster
galaxies. This has been recently investigated by Thomas (1999), from
whom I reproduce here Fig. 5. This shows that the bright galaxies
(those with $L\sim L^*$) produce the bulk of the cluster light, while
the dwarf contribute a negligible amount of light in spite of
dominating the galaxy counts by a large margin. Hence, most galaxies
don't do much, while the brightest $\sim 3\%$ of all galaxies contribute
$\sim 97\%$ of the whole cluster light. Given that it is most
likely that the metals were produced by the same stellar population
that now shines, one can safely conclude that also the bulk of the
cluster metals have been produced by the giant galaxies (or  -- paying a 
tribute to a widespread belief -- 
 by the stars that are now in the giants). Dwarfs 
contribution to ICM metals may have been somewhat larger than their
tiny  contribution to the cluster light, since metals can more
easily escape from their shallower potential wells (Thomas 1999), but this
cannot alter the conclusion that the giants dominate metal production
by a very large factor.

\section{A Semiempirical Estimate of Preheating}

\medskip

The total amount of iron in clusters represnts a record of the overall
past supernova (SN) activity as well as of the past mass and energy
ejected from cluster galaxies.  The empirical values \IMLR \ was used
to set a constraint on the so-called preheating of the ICM (Renzini
1994), a hot topic indeed at this meeting.  The total SN heating is
given by the kinetic energy released by one SN ($\sim 10^{51}$ erg)
times the number of SNs that have exploded.  It is convenient to
express this energy per unit present optical light $\lb$, i.e.:
\begin{equation}
{E_{\rm SN}\over\lb}=10^{51}\,{ N_{\rm SN}\over\lb}=10^{51}\, \left({M_{\rm Fe}\over\lb}
\right)^{\rm TOT}\,{1\over <\! M_{\rm Fe}\! >}\simeq 10^{50}\quad ({\rm erg}/\lsun), 
\end{equation}
where as total (ICM+galaxies) \IMLR =0.03 $\msun/\lsun$ is adopted, and the
average iron mass release per SN event is assumed to be
$0.3\,\msun$. This is the appropriate average if $\sim 3/4$ of all
iron is made by Type Ia SNs, and $\sim 1/4$ by Type II and other SN
types (see R97). Were all the iron made instead by SNIIs, the
resulting SN energy would go up by perhaps a factor of 3 or 4.
This estimate should
therefore be accurate to within a factor of 2 or 3.

The presence of a large amount of iron in the ICM demonstrates that
matter (and then energy) has been ejected from galaxies.  The kinetic
energy injected into the ICM by galactic winds, again per unit cluster
light, is given by 1/2 the ejected mass times the typical wind
velocity squared, i.e.:
\begin{equation}
{E_{\rm w}\over\lb}={1\over 2} {M_{\rm Fe}^{\rm ICM}\over\lb}
\left<\!{v_{\rm w}^2\over
   Z^{\rm Fe}_{\rm w}}\!\right>\simeq 1.5\times 10^{49}{Z^{\rm Fe}_
   \odot\over Z^{\rm Fe}_{\rm w}}\cdot
   \left({v_{w}\over 500\,{\rm km}\,{\rm s}^{-1}}\right)^2\simeq 10^{49}\quad ({\rm erg}/\lsun), 
\end{equation}
where the empirical \IMLR \ for the ICM has been used, the average
metallicity of the winds $Z^{\rm Fe}_{\rm w}$ is assumed to be twice
solar, and the wind velocity $v_{\rm w}$ cannot be much different from
the escape velocity from individual galaxies, as usual in the case of
thermal winds. Again, this estimate may be regarded as accurate to
within a factor of 2, or so.

A first  inference from these estimates is that $\sim 5-10\%$ of the kinetic energy released by 
SNs  survives as
kinetic energy of galactic winds, thus contributing to the heating of the ICM. 
A roughly similar amount should go
into work to extract the gas from the potential well of individual galaxies, while the bulk 
$\sim 80-90\%$ has to be radiated away locally and does not contribute to the feedback.

The estimated $\sim 10^{49}$ erg/\lsun \ correspond to a ``preheating'' of
$\sim 0.1$ keV per particle, for a typical cluster $M_{\rm ICM}/\lb\simeq
40\;\msun/\lsun$. This is $\sim 20$ times lower than the (AGN induced)
preheating some model require
to fit the cluster $L_{\rm X}-T$ relation (Wu et al. 1999).
Not even if all the SN energy were to go to increase the thermal energy of 
the ICM whould this extreme requirement be met.
Crucial for the evolution of the ICM is however the entropy, hence the
relative timing of 
galactic wind heating and cluster formation (Kaiser 1991;
Cavaliere et al. 1993). The entropy increase associated
to a given amount of preheating depends on the gas 
density when the heating takes place.. 
 Most of star formation
in cluster elliptical galaxies --
and therefore most SN activity and galactic winds -- appears to be 
confined at very early times, i.e., at $z\gsim 3$. It remains to be seen
whether such  early preheating may render 
 $\sim 0.1$ keV per particle sufficient to increase enough
the entropy thus producing the observed $L_{\rm X}-T$ relation. 
Alternative models less demanding than those of Wu et al. in terms of
preheating are presented by Tozzi at this meeting (Tozzi \& Norman 1999).
Finally, if AGN preheating were as high as 2 keV/particle (averaged over the 
whole cluster) it may result in long term suppression of cluster cooling flows,
in analogy with the scenario proposed for intermittent gas flows in ellipticals
(Ciotti \& Ostriker 1997).

\section{Clusters as Fair Samples and the Metallicity of the High-$z$ Universe}

In R97 and in some its subsequent updates (e.g. Renzini 1998, 1999) it
is argued that clusters are representative of the local
universe as a whole, as far as the fraction of baryons already turned
into stars is concerned, hence of the global metallicity of the local
universe ($\sim 1/3$ solar). Moreover, since stars in spheroids formed
at $z\gsim 3$ and spheroids account for at least 1/3 of the present
day total mass in stars, it is concluded that the metallicity of the
whole universe at $z\sim 3$ had to be $\sim 1/3$ of the present global
value, hence $\sim 1/10$ solar. This simple argument supports the
notion of a {\it prompt initial enrichment} of the early universe.

\newpage
\noindent
\acknowledgements{I am grateful to Silvia Pellegrini and Daniel Thomas for their permission
to reproduce here Fig. 4 and Fig. 5, respectively.}

\vfill
\end{document}